\documentclass[graybox]{svmult}

\usepackage{mathptmx}       % selects Times Roman as basic font
\usepackage{helvet}         % selects Helvetica as sans-serif font
\usepackage{courier}        % selects Courier as typewriter font
\usepackage{type1cm}        % activate if the above 3 fonts are
\usepackage{makeidx}         % allows index generation
\usepackage{graphicx}        % standard LaTeX graphics tool
\usepackage{multicol}        % used for the two-column index
\usepackage[bottom]{footmisc}% places footnotes at page bottom
\usepackage{cite}
\usepackage{dsfont}

\usepackage{soul}
\setstcolor{red}
\sethlcolor{yellow}
\makeindex             % used for the subject index
\usepackage{amsmath,amssymb,bm}

\newcommand{\bra}[1]{\left<{#1}\right|}
\newcommand{\ket}[1]{\left|{#1}\right>}
\newcommand{\braket}[2]{\left<\left.{#1}\right|{#2}\right>}
\newcommand{\ave}[1]{\left\langle #1\right\rangle}
\newcommand{\wv}[3]{\vphantom{\langle #2\rangle}_{#1}\langle #2 \rangle_{#3}}

\newcommand{\re}{\rm Re}

\begin{document}

\title*{Standard and Null Weak Values}
% Use \titlerunning{Short Title} for an abbreviated version of
% your contribution title if the original one is too long
\author{Oded Zilberberg, Alessandro Romito and Yuval Gefen}
% Use \authorrunning{Short Title} for an abbreviated version of
% your contribution title if the original one is too long
\institute{Oded Zilberberg \at Department of Condensed Matter Physics, Weizamnn Institute of Science, Rehovot 76100, Israel, \email{oded.zilberberg@weizmann.ac.il}
\and Alessandro Romito \at Dahlem Center for Complex Quantum Systems and Fachbereich Physik, Freie Universit\"at Berlin, 14195 Berlin, Germany, \email{alessandro.romito@fu-berlin.de} \and Yuval Gefen \at Department of Condensed Matter Physics, Weizamnn Institute of Science, Rehovot 76100, Israel, \email{yuval.gefen@weizmann.ac.il}}
%
% Use the package "url.sty" to avoid
% problems with special characters
% used in your e-mail or web address
%
\maketitle

\abstract{Weak value (WV) is a quantum mechanical measurement protocol, proposed by Aharonov, Albert, and Vaidman. It consists of a weak measurement, which is weighed in, conditional on the outcome of a later, strong measurement. Here we define another two-step measurement protocol, null weak value (NVW), and point out its advantages as compared to WV. We present two alternative derivations of NWVs and compare them to the corresponding derivations of WVs.}

\section{Introduction}
This contribution is dedicated to Yakir Aharonov, on his $80^{\rm th}$ birthday. His seminal work in quantum mechanics, and the stimulating discussions we have had with him, have influenced our own work in physics in a deep way. YG is indebted to Yakir Aharonov for the close interaction, and his continuous support over the years.

The von Neumann formulation of measurement in quantum mechanics invokes a generic Hamiltonian of the form\cite{von-neumann},
\begin{equation}
\label{ham}
H=H_{\textrm{S}}+H_{\textrm{D}}+H_{\textrm{int}} \, ,
\end{equation}
where $H_{\textrm{S}}$ is the Hamiltonian of the system to be measured, $H_{\textrm{D}}$ is the detector's Hamiltonian, and $H_{\textrm{int}}$ represents the coupling between the two:
\begin{align}
H_{\textrm{int}}=\lambda\, g(t) \hat{p}\hat{A} \, .	
\label{interact}
\end{align}
Here $\hat{p}$ is the momentum canonically conjugate to the position of the detector's pointer, $\hat{q}$, $g(t)$ represents the time window during which the measurement (system-detector coupling) takes place, and $\lambda$ is the dimensionless strength of the measurement. The measured observable is 
$\hat{A}=\sum_i a_i \ket{a_i}\bra{a_i}$.

\begin{figure}
\begin{center}
\includegraphics[width=70mm]{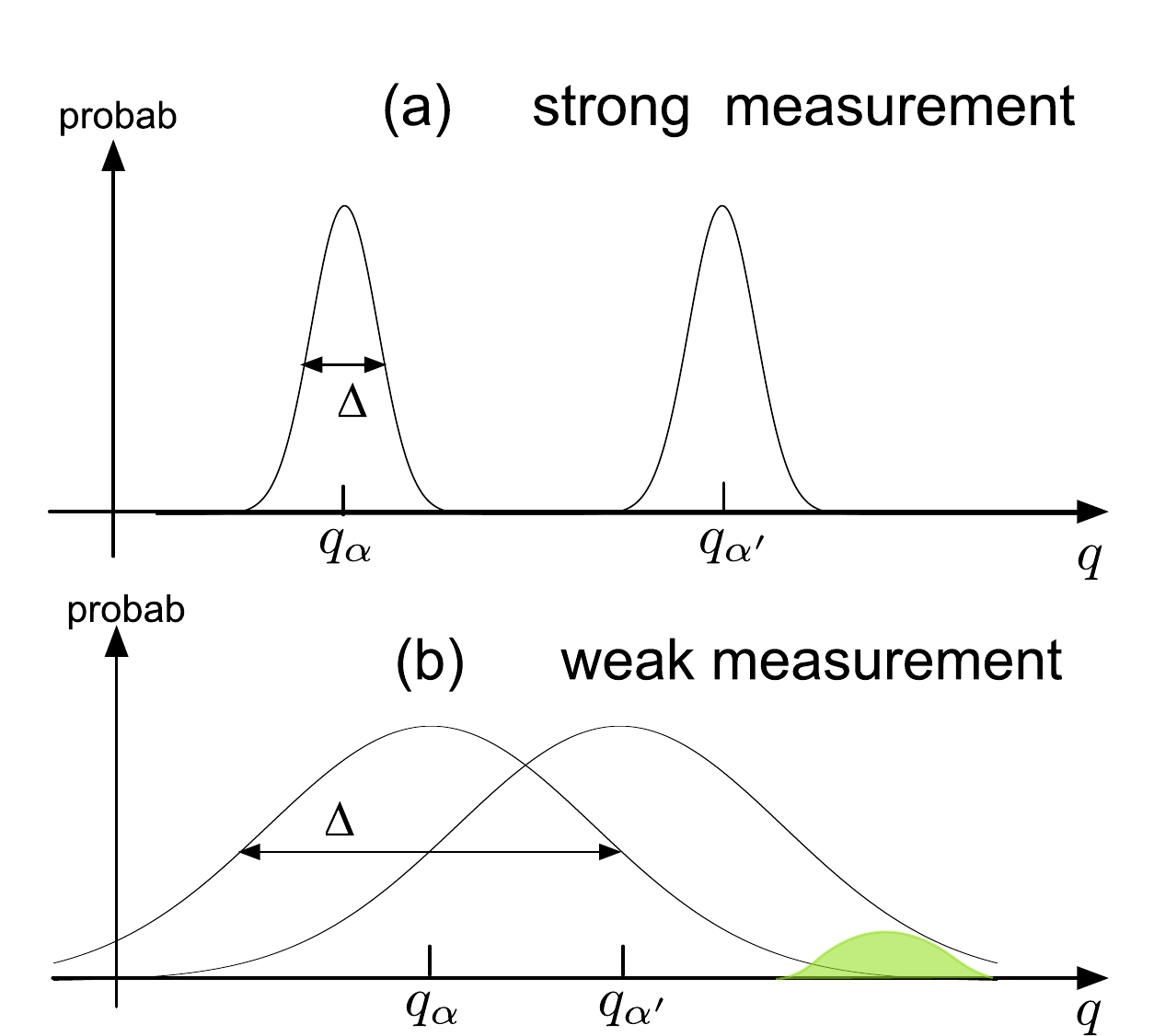}
\end{center}
\caption{Strong and weak measurements. When the system is in a
given eigenstate, $\ket{\alpha}$, of the measured observable,
$\hat{A}$, the coordinate shift in the detector, $q$, has a
certain probability distribution, $P_{\alpha}(q)$. (a) When such
distributions are well peaked around different values of
$q_{\alpha}$, i.e. $\Delta \ll |q_{\alpha'}-q_{\alpha}|$, one can
uniquely associate a value of the coordinate, $q_{\alpha}$, to
each eigenstate of $\hat{A}$, $\ket{\alpha}$, which corresponds to
a strong measurement. (b) In the opposite case of a weak
measurement, $\Delta \gg |q_{\alpha'}-q_{\alpha}|$, the small
shift in the detector's coordinate (due to the weak
system-detector coupling) is blurred by the distribution
uncertainty. The shaded (green) area on the right represents the part of the distribution selected by the WV protocol.}
\label{fig1}
\end{figure}

Strong measurement is associated with the collapse of the wave function dogma \cite{von-neumann}. In an ideal strong measurement there is a one-to-one correspondence between the observed value of the detector's coordinate,  $q_{\alpha}$, and the eigenstates $\ket{\alpha}$ of the system's measured operator, $\hat{A}$ [cf.~Fig.~\ref{fig1}(a)]. In a weak measurement ($\lambda \ll 1$), instead, the ranges of values of  $q$  that correspond
 to two distinct eigenstates of $\hat{A}$, $\ket{\alpha}$  and $\ket{\alpha'}$,  are described by two strongly overlapping probability distribution functions,
$P_{\alpha}(q)$  and $P_{\alpha'}(q)$, respectively.  Hence the measurement of $q$
 provides only partial information on the state of S [cf.~Fig.~\ref{fig1}(b)] and changes its state only slightly. Nonetheless the mean value of a single weak measurement (averaged over many repetition of the measurement) coincides with that of a strong measurement.

One may extend the notion of weak measurement by referring to a sequence of correlated (especially conditional) measurements. Conditional quantum measurements can lead to results that
cannot be interpreted in terms of classical probabilities, due to the
quantum correlations between measurements.
An intriguing example for correlated quantum measurements outcome is the so called \emph{weak value} (WV). It is the outcome of a measurement scheme originally developed by Aharonov, Albert and Vaidman~\cite{Aharonov:1988aa}.
The WV measurement protocol consists of (i) initializing the system in
a certain pure state  $|\, i\, \rangle$ (preselection; generalization to a mixed state is possible \cite{Romito:2010} but will not be discussed here), (ii) weakly measuring observable $\hat{A}$ of the system,
(iii) retaining the detector output only if the system is eventually
measured to be in a chosen final state $|\, \textrm{f}\,
\rangle$ (postselection).
The average signal monitored by the detector will then be proportional
to the real (or imaginary) part of the complex WV,
\begin{align}
\label{eq:weak} 
\wv{f}{\hat{A}}{i} = \frac{\langle\, \textrm{f} \,| \,\hat{A}\,
| i\, \rangle   }{   \langle \,\textrm{f}\, |  i \rangle}\, ,
\end{align}
rather than to the standard average value, $\bra{i} \hat{A} \ket{i}$ [cf.~Fig.~\ref{fig1}(b)]. Further discussion of the context in which WV should be understood has been provided~\cite{Wiseman:2002,Jozsa:2007,Dressel:2010}.

Going beyond the peculiarities of WV protocols,  recent series of works explored the potential of WVs in quantum optics~\cite{Ritchie:1990,Pryde:2005,Hosten:2008,Dixon:2009,Starling:2009,Brunner:2010,Starling:2010b} and solid-state physics~\cite{Williams:2008,Romito:2008,Shpitalnik:2008,Zilberberg:2011}, ranging from experimental observation to their application to hypersensitive measurements. In the latter, a measurement, performed by a detector \emph{entangled} with a system whose states can be preselected and postselected, leads to an amplified signal in the detector that enables sensing of small quantities~\cite{Hosten:2008,Dixon:2009,Starling:2009,Brunner:2010,Starling:2010b,Zilberberg:2011}.
Quite generally, within a WV-amplification protocol, only a subset of the detector's readings, associated with the tail of the signal's distribution, is accounted for. Notwithstanding the scarcity of data points, the large value of ${}_f\langle \hat{A}\rangle_i$, leads to an amplification~\cite{Starling:2009,Zilberberg:2011} of the signal-to-noise ratio (SNR) for systems where the noise is dominated by an external (technical) component.

The amplification originating from WV protocols is non-universal. The specifics of such amplification are diverse and system-dependent.	In fact, for statistical (inherent) noise,  SNR amplification resulting from  large WVs is generally suppressed due to  the reduction in the statistics of the collected data:  postselection restricts us to a small subset of the readings at the detector.  The upside of the WV procedure has several facets: if we try to enhance the statistics  by increasing the intensity of input signal through the system (e.g., intensity of the impinging photon beam), possibly entering a non-linear response regime, postselection will effectively reduce this intensity back to a level accessible to the detector sensitivity~\cite{Dixon:2009}. Alternatively, amplification may originate  from the imaginary component of the WV~\cite{Hosten:2008},  or from the different effect of the noise and the measured variable on the detector's signal~\cite{Zilberberg:2011}.
 However, as long as quantum fluctuations (leading to inherent statistical noise)
dominate, the large WV is outweighed by the scarcity of data points, failing to amplify the signal-to-statistical-noise ratio~\cite{Zilberberg:2011,Zhu:11}.

We have recently  presented an alternative measurement protocol dubbed \textit{null weak value} (NWV), that overcomes the SNR problem of WV-protocols~\cite{Zilberberg:12}. Like the WV-protocol, NWV consists of a two-step conditional measurement. The recipe goes as follows: (i) We prepare the system in a given pure state $\ket{i}$ (a generalization to a mixed state will not be discussed here). (ii) We perform a strong measurement of the observable $\hat{A}$; we arrange the setup such that the probability, $p$, that our detector ``clicks'' (hence collapse is taking place), is small ($p \ll 1$). If the detector clicks, the system's state has collapsed, and we start our measurement all over again, with a new replica of the system. If no click has occurred, $\ket{i}$ has been (by way of back-action) modified, $\ket{i}\rightarrow\ket{i_p}$. (iii) We now let the system evolve in time, possibly manipulating $\ket{i_p}$ in a controlled way. (iv) We perform a strong measurement of a certain observable $\hat{B}$. Formulating the results of the first and second measurements in an anti-causal manner, the outcome of the first measurement (the detector clicking) is conditional on the outcome of the second measurement (of $\hat{B}$). The NWV of $\hat{A}$ is then 
\begin{align}
\label{qm_nwv_def}
\ave{\hat{A}}_{\rm NWV} = \frac{\langle i|\hat{A}|i\rangle} {|\langle f|i\rangle|^2} \, ,
\end{align}
to be compared with the WV of $\hat{A}$, Eq.~\eqref{eq:weak}.

In principle, the derivation of standard WVs, as well as that of NWVs, can be done by following the dynamics of the measurement process in the extended system-detector Hilbert space. Here we discuss the derivation of NWVs and compare it to the derivation of WVs, taking two different paths: (i) analyzing the the effect of the detector on states and amplitudes in the system's subspace. We refer to this as ``derivation in terms of quantum  states''. (ii) We derive expressions for WVs / NWVs analyzing the probabilities and conditional probabilities involved in the various steps of the protocols.

The outline of this paper is as follows. In Section \ref{std_wv} we present a derivation of a standard WV in terms of quantum states. This follows by a derivation in Section \ref{cond}, which employs conditional probabilities. Section \ref{nwv_cond} addresses NWV from the viewpoint of conditional probabilities, and Section \ref{state} presents a derivation of NWV in terms of quantum states. In Section \ref{conclusion} we outline a few conclusions.

\section{Standard WV Derivation}
\label{std_wv}
Weak values describe the outcome read in a detector when the measured system is subsequently found to be in a specific state. 
The expression for WVs can be derived most simply through an argument due to Yakir Aharonov based on a one-line identity for the average value of an observable $\hat{A}$:
\begin{align}
\label{eq:weak2}
\ave{\hat{A}} &= \bra{i}\hat{A}\ket{i}=\sum_n \braket{i}{f_n}\bra{f_n}\hat{A}\ket{i}=\sum_n |\braket{f_n}{i}|^2 \frac{\bra{f_n}\hat{A}\ket{i}}{\braket{f_n}{i}}\equiv\sum_n P_{i\rightarrow n}\;\wv{f_n}{\hat{A}}{i}\, .
\end{align}
The identity is obtained by inserting a complete set of states $1 = \sum_n\ket{f_n}\bra{f_n}$. Also in the last equality we introduced the notation $\wv{f_n}{\hat{A}}{i} \equiv \bra{f_n}\hat{A}\ket{i}/\braket{f_n}{i}$, $P_{i\rightarrow n} \equiv |\braket{f_n}{i}|^2$. 
The reasoning goes as follows:
The states $\left\{\ket{f_n}\right\}$ above can be interpreted as the possible states obtained by measuring an observable $\hat{B}$ after $\hat{A}$.  If one can assume that the measurement $\hat{A}$ \emph{leaves  the initial state unchanged}, $P_{i\rightarrow n}$ can be interpreted as the probability that the system is (finally) to be found in  $\ket{f_n}$.
Eq.~(\ref{eq:weak}) gives then a natural interpretation of the WV, $\wv{f_n}{\hat{A}}{i}$, as the average of $\hat{A}$ conditional to a postselection on $\ket{f_n}$.
The so obtained expression for the WV is universal in the sense that it does not depend on the the specific detector or its coupling to the system, as long as the state of the system is unaffected by the detection process. 
Not modifying the state of the system has to do with the measurement back-action and its precise meaning is what defines the weakness of the measurement, hence the name weak value. 
The weakness of the measurement can be addressed in a specific model of the system-detector coupling. One may, of course, reproduce the correct expression for the WV in a treatment involving the system-detector Hilbert space.

The weak coupling between a system and a detector is performed by an ideal von Neumann measurement~\cite{von-neumann}, described by the Hamiltonian in Eqs.~\eqref{ham} and \eqref{interact} with $\lambda \ll 1$. We assume for simplicity that the free Hamiltonians of the system and the detector vanish and that $g(t)=\delta(t-t_0)$.

The system is initially prepared in the state $\ket{i}$, and the detector in the state $\ket{\phi_0}$. The latter is assumed to be a Gaussian wave-packet centered at $q=q_0$, $\ket{\phi_0}=C e^{-(q-q_0)^2/4\Delta^2}$.
After the interaction with the detector the entangled state
of the two is
\begin{equation}
\label{stato_entangled}
\ket{\psi}=e^{-i\lambda \hat{p} \hat{A} } \ket{i}\ket{\phi_0} \, .
\end{equation}
In a regular measurement the signal in the detector, i.e. the pointer's position $\langle q \rangle= q_0 +\lambda \langle \hat{A} \rangle$, is read.
From the classical signal, $\langle q \rangle$, one can infer the average value of the observable $\hat{A}$.

In a WV protocol the signal in the detector is kept provided that the system is successfully postselected to be in a state $\ket{f}$.
Hence, the detector ends up in the state
\begin{align}
\label{spostamento}
\ket{\psi}&=\ket{\phi_0} -i \lambda \left[\langle f|\hat{A}|i \rangle/
\langle f|i \rangle\right] \hat{p}\ket{\phi_0}\nonumber\\
&\approx
e^{-i \lambda {}_f\langle\hat{A}\rangle_i} \hat{p}
\ket{\phi_0} \, ,
\end{align}
that corresponds to a shift in the position of the pointer proportional to $ \mathop{Re}[_{f} \langle A \rangle_{i} ]$.
Hence the expectation value of the coordinate of the pointer is given by
\begin{equation}
{}_f\langle \hat{q}\rangle_i=q_0+\lambda  \mathop{Re}[_{f} \langle \hat{A} \rangle_{i} ]\, ,
\end{equation}
where the conditional average value of $\hat{A}$ is inferred from the detector's reading.

We note that the approximation in Eq.~\ref{spostamento} is valid when $\Delta \gg \max_{i,j} |a_i-a_j|$.
This means that the initial detector's wave function and the shifted one due to the interaction with the system are strongly overlapping. In turn, this means that for \emph{any} outcome of the detector the state of the system is weakly changed. This corresponds to a weak measurement. As long as the measurement of the observable, $\hat{A}$, is weak, the WV is system independent and does not depend on the details of the coupling or the specific choice of the detector.

\section{Standard Weak Values in Terms of Conditional Probabilities}
\label{cond}
In this section we provide a correspondence between a conditional probability notation and the standard state-vector notation used in most WV works.

Let us consider the case of a qubit system weakly coupled to another two-level detector. Their respective states are initially
\begin{align}
\ket{i} = \alpha \ket{0} + \beta \ket{1} &\quad\quad\quad \text{Hilbert space of system}\, ,\\
\ket{d} = \gamma \ket{0} + \delta \ket{1}&\quad\quad\quad \text{Hilbert space of detector}\,.
\end{align}
Once the system and detector are coupled their resulting entangled state is
\begin{align}
\ket{i,d} =\mathcal{N}  \alpha\gamma \ket{0,0} + \alpha\delta \ket{0,1} + \beta \tilde{\gamma} \ket{1,0} + \beta\tilde{\delta} \ket{1,1}\, ,
\end{align}
where we assumed that if the system is in state $\ket{0}$ the detector remains unaffected, $\tilde{\gamma},\tilde{\delta}$ represent the detector amplitudes when the system is in state $\ket{1}$, and $\mathcal{N}$ is a normalization factor.

Measuring the detector state and applying calibration yields an observable of the system:
\begin{align}
\bra{ i}\hat{A}\ket{i} = \frac{P( \hat{A}') - |\delta|^2}{|\tilde{\delta}|^2 - |\delta|^2} = |\beta|^2\, ,
\end{align}
where $\hat{A}=\ket{1}\bra{1}$ operates in the system's Hilbert space and $\hat{A}'=\ket{0,1}\bra{0,1}+\ket{1,1}\bra{1,1}$ operates in the joint Hilbert space of system and detector.

\begin{figure}
\begin{center}
\includegraphics[width=\textwidth]{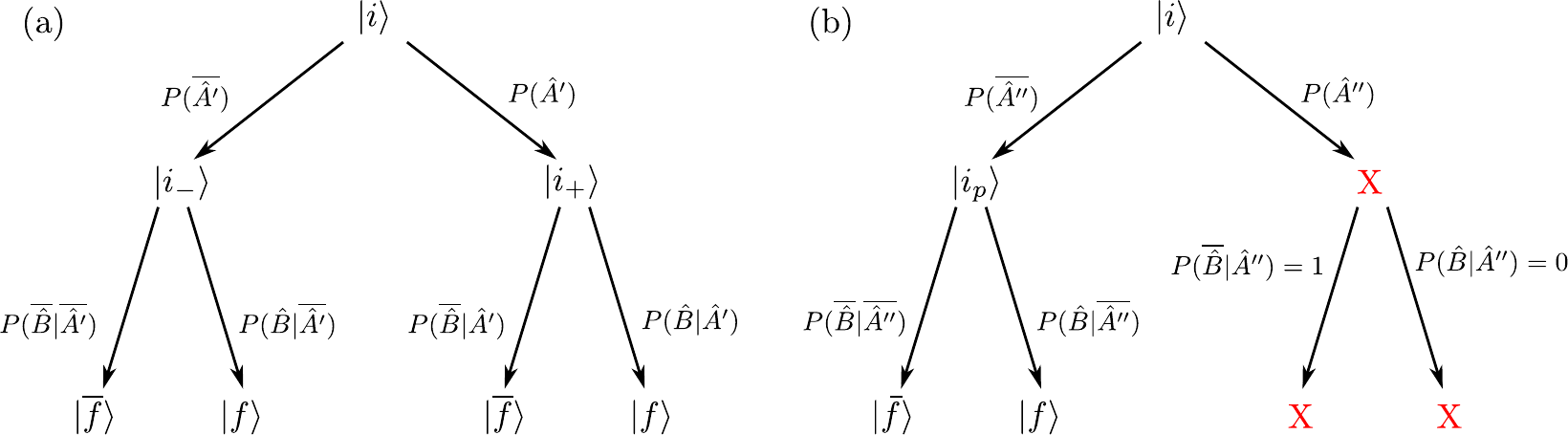}
\end{center}
\caption{Tree diagrams  of the state evolution during the (a) weak value protocol and (b) null weak value protocol. The respective  probabilities of events are indicated. (a) Upon the first weak measurement the detector ``clicks'' [no ``click''] with probability $P(\hat{A}')$ [$P(\overline{\hat{A}'})$]. Due to the measurement back-action, the system evolves into the state $\ket{i_+}$ [$\ket{i_-}$]. Subsequently the system if strongly measured to be in the $\ket{f}$ [or not]. (b) The first partial-collapse measurement ``clicks'' [no ``click''] with probability $P(\hat{\tilde{A}})$ [$P(\overline{\hat{\tilde{A}}})$]. If a ``click'' occurred the system is destroyed [marked by $X$ (red)]. If not, due to the measurement back-action, the system evolves into the state $\ket{i_p}$. Subsequently the system if strongly measured to be in the $\ket{f}$ [or not].}
\label{fig2}
\end{figure}

Taking the probability of this weak measurement outcome conditional on the outcome of a subsequent postselection, yields the detector response to a standard WV protocol
\begin{align}
\frac{P(\hat{A}'\, |\,\hat{B}) - |\delta|^2}{|\tilde{\delta}|^2 - |\delta|^2} \sim (|\tilde{\delta}|^2 - |\delta|^2)\re\left\{\wv{f}{\hat{A}}{i}\right\}\,,
\end{align}
where $\hat{B}=\ket{f}\bra{f}$ is a strong postselection in the system space, and we used the weakness of the first measurement $|\tilde{\delta}|^2 - |\delta|^2 << 1$. The counter-causal conditional probability $P(\hat{A}'\, |\,\hat{B})$ is calculated using Bayes theorem and using the causal conditional probabilities appearing in the tree diagram in Fig.~\ref{fig2}(a).

%%%%%%%%%%%%%%%%%%%%%%%%%%%%%%%%%%%%%%%%%%%%%%%%%
%%%%%%%%%%%%%%%%%%%%%%%%%%%%%%%%%%%%%%%%%%%%%%%%%%%%

\section{Null Weak Values in Terms of Conditional Probabilities}
\label{nwv_cond}
We now turn to describe our new measurement protocol (null-WV) [cf. Fig.\ref{fig2}(b)]. The qubit state is measured twice.
The first measurement $\hat{\tilde{A}}=p_0 \ket{0}\bra{0}+p_1\ket{1}\bra{1}$ is a strong (projective) measurement which is performed on the system with small probability.
Here the states $\{\ket{0},\ket{1}\}$ are measured with probabilities $\{p_0,p_1\}$, respectively.
For simplicity, hereafter, we assume that only the state $\ket{1}$ is measured with probability $p_1=p$ and $p_0=0$.
If the detector ``clicks'' (the measurement outcome is positive), the qubit state is destroyed. Very importantly, having a ``null outcome'' (no click) still results in a weak back-action on the system. We refer to this stage of the measurement process as ``weak partial-collapse''.
Subsequently the qubit state is (strongly) measured a second time (postselected), $\hat{B}$, to be in the state $\ket{f}$ (click) or $\ket{\bar{f}}$ (no click). 

Similarly to the previous case, readout of the number of ``clicks'' in this first detector yields an observable on the system
\begin{align}
\bra{ i}\hat{A}\ket{i} = \frac{P( \hat{\tilde{A}} )}{p} = |\beta|^2\,.
\end{align}
Studying the correlation between this first partial-collapse measurement and a ``no click'' postselection yields the null value
\begin{align}
\frac{P( \hat{\tilde{A}} | \overline{\hat{B}} )}{p} &= \frac{\bra{ i}\hat{A}\ket{i}}{ P( \overline{\hat{B}})} \sim \frac{\bra{ i}\hat{A}\ket{i}}{ |< f | i >|^2}\,,
\label{nwv}
\end{align}
where the last approximation is for small $p$.

Our protocol takes advantage of the correlation between the two measurements. To shed some light on its outcome we calculate explicitly the conditional probabilities following the measurement procedure sketched in Fig.~\ref{fig2}(b).  For example, if the first measurement results in a ``click'' the system's state is destroyed and any subsequent measurement on the system results in a null-result, implying $P(\hat{B}|\hat{\tilde{A}})=0$, and $P(\overline{\hat{B}}|\hat{\tilde{A}})=1$.
This represents a classical correlation between the two measurements. By contrast, $P(\overline{\hat{B}}|\overline{\hat{\tilde{A}}})$ embeds non-trivial quantum correlations.
The first partial-collapse measurement of a given preselected state $\ket{i}$ results in the detector clicking with probability $P(\hat{\tilde{A}})=p|\beta|^2$.
If no click occurs [with probability $P(\overline{\hat{\tilde{A}}})=1-P(\hat{\tilde{A}})$], the qubit's state is modified by the measurement back-action into $\ket{i_{p}}=\big[\alpha\ket{0}+\sqrt{1-p}\beta\ket{1}\big]/\sqrt{P(\overline{\hat{\tilde{A}}})}$.
The second strong measurement, $\hat{B}$, yields a click [no click] with probability $P(\hat{B}|\overline{\hat{\tilde{A}}})=|\langle f|i_{p}\rangle |^2$   $\left[P(\overline{\hat{B}}|\overline{\hat{\tilde{A}}})=|\langle \bar{f}|i_{p}\rangle |^2\right]$.
Finally, using Bayes theorem, we can write $ P(\hat{\tilde{A}}|\overline{\hat{B}}) = P(\hat{\tilde{A}}) / [ P(\hat{\tilde{A}}) + P(\overline{\hat{\tilde{A}}}) P(\overline{\hat{B}}|\overline{\hat{\tilde{A}}}) ]$. This correlated outcome is useful in obtaining amplified SNR in a quantum state discrimination problem \cite{Zilberberg:12}.

\section{Description of Standard and Null Weak Values in Terms of Quantum States}
\label{state}

The result of the NWV protocol, Eq.~(\ref{qm_nwv_def}), though emanating from a weak measurement, is different from the standard WV, Eq.~(\ref{eq:weak}).
On the face of it, the derivation that leads to Eq.~(\ref{eq:weak2}) appears to be universally adapted to any two-step (the first is weak) measurement procedure. The fact that the expression for NWV is different from that of WV may then seem paradoxical. 
It is therefore instructive to understand how the NWV relates to the standard WV in the framework of the derivation of Eq.~(\ref{eq:weak2}).

The idea behind Eq.~(\ref{eq:weak2}) is quite general: one writes an identity for the standard quantum mechanical expectation value in terms of a sum of probabilities to reach the possible postselected states. These probabilities are weighted-in with the appropriate coefficients, namely
\begin{equation}
\ave{\hat{A}} = \sum_n P_{i\rightarrow n}\; \wv{f_n}{\hat{A}}{i}\, ,
\label{template}
\end{equation}
where $P_{i\rightarrow n}$ is the probability to obtain the postselected state
$\ket{f_n}$ given the initial (prepared) state of the system is unchanged. The coefficients $\wv{f_n}{\hat{A}}{i}$ obtained from this identity are then naturally interpreted as the conditional averages one is after. 
In the (standard WV) derivation of Eq.~(\ref{eq:weak2}), the first weak measurement does not affect the state significantly, and this interpretation comes natural with $P_{i\rightarrow n}=|\langle f_n|i\rangle|^2$.
The factor $|\langle f_n|i\rangle|^2$
can no longer be interpreted as the probability to find the system
in $\ket{f_n}$, following the second strong measurement (postselection).
In fact, (without extending our Hilbert space) the partial-collapse measurement cannot be directly described by a von
Neumann-like measurement on which Eq.~(\ref{eq:weak2}) is implicitly
based. 
Therefore, despite the fact that Eq.~(\ref{eq:weak2}) is an identity, which holds true also in the NWV case, it does not allow for an interpretation of the conditional outcome as the NWV.

Nevertheless, it is possible to write other identities in the spirit of Eq.~(\ref{template}) which can be useful in the present case. 
For example, following the steps (i) using $\hat{A}=\sum_j a_j \hat{\Pi}_j$, where $\left\{a_j\right\}$ are  the eigenvalues of $\hat{A}$ and $\left\{\hat{\Pi}_j\right\}$ the corresponding projection operators onto the states $\left\{\ket{j}\right\}$, (ii) inserting the projector identity $\hat{\Pi}_j\equiv \hat{\Pi}_j^2$, and (iii) inserting the identity in terms of postselected states, $\mathds{1}=\sum_n \ket{f_n}\bra{f_n}$, one obtains
\begin{align}
\ave{\hat{A}} &= \sum_n \left( \sum_j |\ave{f_n|\Pi_j|i}|^2 \right) \frac{\sum_j a_j |\ave{f_n|\Pi_j|i}|^2}{\sum_j |\ave{f_n|\Pi_j|i}|^2}\, .
\label{strong}
\end{align}
If the measurement of $\hat{A}$ is strong, the term in parenthesis in
Eq.~\eqref{strong} corresponds to the probability of
postselection. Therefore the conditional average is the remaining
expression outside of the parenthesis. Equation \eqref{strong} is evidently an identity, but it does not lend itself to any physically meaningful interpretation if the protocol involves weak measurement.

Since partial-collapse measurements are in fact strong measurements that occur with a small probability, Eq.~\eqref{strong} is particularly useful for NWVs.
Indeed, one may effectively describe a partial-collapse measurement as a von Neumann measurement in an extended Hilbert space. 
To do so, we formally extend the system's Hilbert space to include an
extra ancilla state, $\ket{R}$.
The idea is to describe the partial-collapse measurement as a combination of a weak transition to the ancilla state followed by a strong projective
measurement exactly of this newly added state. 
Let us describe this more precisely.

% We have seen that both the standard and the null WV measurement protocol can be described in terms of conditional probabilities and in terms of state vectors.
% Let us see how they behave within's Aharonov's frameword of writing a mean observable in terms of WVs:
% \begin{align}
% \label{weak}
% \ave{A} &= \bra{i}\hat{A}\ket{i}=\sum_n \braket{i}{f_n}\bra{f_n}\hat{A}\ket{i}=\sum_n \braket{f_n}{i}|^2 \frac{\bra{f_n}\hat{A}\ket{i}}{\braket{f_n}{i}}=\sum_n P_{i\rightarrow n}\wv{f_n}{\hat{A}}{i}\, .
% \end{align}

The initial state one is interested in is $\ket{i}=\alpha \ket{0}+\beta
\ket{1}$. Allowing the state $\ket{1}$ to be transferred to $\ket{R}$ with transition rate $\Gamma$ for a time window $t$, evolves the initial state into $\hat{U}\ket{i} = \alpha \ket{0}+\sqrt{1-p}\beta\ket{1} + \sqrt{p}\beta \ket{R}$, where $p=1-\exp(-\Gamma t)$ is the probability to undergo this transition over time $t$, and $\hat{U}\ket{i}$ is in the extended Hilbert space spanned by $\ket{0}, \ket{1},\ket{R}$.  It is now apparent that the partial-collapse measurement can be written as $\hat{\tilde{A}} =
\hat{U}^\dagger\ket{R}\bra{R}\hat{U}$. Subsequently we postselect on 
$\ket{f}$ (a state within the system's Hilbert space, $\ket{0}, \ket{1}$). Hence, in the extended Hilbert space the measurement can be formulated
according to the standard measurement theory.

Let us adjust Eq.~\eqref{strong} to the NWV case,
\begin{align}
\label{strongNWV}
\ave{\hat{A}}&=\frac{\ave{\hat{\tilde{A}}}}{p} \\
&= \frac{1}{p}\sum_{f_n=f,\bar{f}} \left( \sum_{j=\bar{R},R} |\ave{f_n|\Pi_j\hat{U}|i}|^2 \right) \frac{ |\ave{f_n|\Pi_R\hat{U}|i}|^2}{\sum_{j=\bar{R},R} |\ave{f_n|\Pi_j\hat{U}|i}|^2}\quad \text{(inappropriate!)}\, ,\nonumber
\end{align}
where $\Pi_{\bar{R}}=\ket{0}\bra{0}+\ket{1}\bra{1}$ is the projector onto the Hilbert space of the original
system, i.e. the subspace orthogonal to $\ket{R}$. Owing to the fact that the postselection states $\ket{f}$ and $\ket{\bar{f}}$ are within the reduced Hilbert space, the numerator on the right hand side is identically zero. 

In order to harness this approach to describe the procedure at hand, we introduce a crucial modification in our scheme. One may think of the postselection as performing an additional
``partial-collapse'' measurement with the state $\ket{f}$ having probability $p\equiv 1$ to become transmitted into state $\ket{R}$, i.e. the postselection is with respect to the operator (observable) $\hat{\tilde{B}}=
\hat{\tilde{U}}^\dagger\ket{R}\bra{R}\hat{\tilde{U}}$, with $\hat{\tilde{U}}\left[a\ket{\bar{f}}+b\ket{f}\right]=\left[a\ket{\bar{f}}+b\ket{R}\right]$. Due to the specific nature of the partial-collapse measurement, the state $\ket{R}$ is not affected by
this transformation. In particular, it does \emph{not} couple back to the
original Hilbert space during the second tunneling event. Note that this renders
the evolution \emph{non-unitary} in the extended Hilbert space. This prescription yields,
%  If the operation were trully unitary, during the tunneling \emph{the particle on the
% right should also tunnel back to the reduced Hilbert space}. But this is not
% the case, once the particle is on the right, it is gone. 
\begin{align}
\ave{\hat{A}} &= \frac{1}{p}\sum_{f_n=R,\bar{f}} \left( \sum_{j=\bar{R},R} |\ave{f_n|\hat{\tilde{U}}\Pi_j\hat{U}|i}|^2 \right) \frac{ |\ave{f_n|\hat{\tilde{U}}\Pi_R\hat{U}|i}|^2}{\sum_{j=\bar{R},R} |\ave{f_n|\hat{\tilde{U}}\Pi_j\hat{U}|i}|^2}\nonumber\\
&\equiv P_{i\rightarrow f} \frac{|\bra{R}\hat{U}\ket{i}|^2}{p\;P_{i\rightarrow f}} \quad\quad \text{(appropriate for NWV)}\, ,
\label{strongNWV2}
\end{align}
with $P_{i\rightarrow f}\equiv|\ave{R|\hat{\tilde{U}}\Pi_{\bar{R}}\hat{U}|i}|^2 + |\ave{R|\hat{U}|i}|^2$. Indeed, the expression on the left hand side corresponds to the probability to end up in state ``$\ket{f}$''$\equiv \hat{\tilde{U}}^\dagger \ket{R}$. Note that for $p \ll 1$,  the probability $P_{i\rightarrow f}$ reduces to the form $|\braket{f}{i}|^2$, the same as in the standard WV case [cf.~Eq.~(\ref{eq:weak})]. Last but not least, the multiplicative term in the middle equality of Eq.~\eqref{strongNWV2} (multiplying the parenthesis), is identical to the expression for the NWV [r.h.s.~of Eq.~\eqref{nwv}]. Hence, we have cast the NWV in the form of Eq.~\eqref{template}.

\section{Conclusions}
\label{conclusion}
We have presented here a novel measurement protocol. Similarly to standard weak values, the outcome of this protocol -- null weak value -- is the result of a first (weaker) measurement correlated with a strong postselection. Ostensibly, as long as a single measurement is concerned, the first measurement in both protocols yields the same outcome. However, the substantial difference between the standard- and null- WVs comes to show that back-action on the system is profoundly different. Hence, involving a postselection leads to qualitatively different correlated results.

\begin{acknowledgement}
This work was supported by GIF, the Israel science foundation, Minerva Foundation of the DFG, Israel-Korea MOST grant, and EU GEOMDISS.
\end{acknowledgement}

%\bibliographystyle{spphys}
%\bibliography{weak_values}

\end{document}